\documentstyle[12pt]{article}

\font\twlgot =eufm10 scaled \magstep1
\font\egtgot =eufm8
\font\sevgot =eufm7

\font\twlmsb =msbm10 scaled \magstep1
\font\egtmsb =msbm8
\font\sevmsb =msbm7

\newfam\gotfam
\def\pgot{\fam\gotfam\twlgot}
\textfont\gotfam\twlgot
\scriptfont\gotfam\egtgot
\scriptscriptfont\gotfam\sevgot
\def\got{\protect\pgot}

\newfam\msbfam
\textfont\msbfam\twlmsb
\scriptfont\msbfam\egtmsb
\scriptscriptfont\msbfam\sevmsb

\def\pBbb{\relax\ifmmode\expandafter\Bb\else\typeout{You cann't use
Bbb in text mode}\fi}
\def\Bb #1{{\fam\msbfam\relax#1}}

\def\thebibliography#1{\bigskip\section*{\large
\bf References\\}\list
  {[\arabic{enumi}]}{\settowidth\labelwidth{#1}\leftmargin\labelwidth
    \advance\leftmargin\labelsep
    \usecounter{enumi}}
    \def\newblock{\hskip .11em plus .33em minus .07em}
    \sloppy\clubpenalty4000\widowpenalty4000
    \sfcode`\.=1000\relax}

\def\op#1{\mathop{{\it\fam0} #1}\limits}

\newcommand{\id}{{\rm Id\,}}

\newcommand{\hm}{{\rm Hom\,}}
\newcommand{\dif}{{\rm Diff\,}}

\newcommand{\beq}{\begin{equation}}
\newcommand{\eeq}{\end{equation}}
\newcommand{\ben}{\begin{eqnarray}}
\newcommand{\een}{\end{eqnarray}}
\newcommand{\be}{\begin{eqnarray*}}
\newcommand{\ee}{\end{eqnarray*}}
\newcommand{\bea}{\begin{eqalph}}
\newcommand{\eea}{\end{eqalph}}

\newcommand{\cA}{{\cal A}}

\newcommand{\gf}{{\got f}}
\newcommand{\gj}{{\got J}}

\newcommand{\cZ}{{\cal Z}}

\newcommand{\cK}{{\cal K}}

\newcommand{\dl}{\delta}

\newcommand{\f}{\phi}
\newcommand{\vf}{\varphi}

\newcommand{\m}{\mu}

\newcommand{\bb}{{\bf 1}}

\newcommand{\dr}{\partial}

\newcommand{\ar}{\op\longrightarrow}

\newcommand{\ot}{\otimes}

\newenvironment{eqalph}{\stepcounter{equation}
\setcounter{equationa}{\value{equation}}
\setcounter{equation}{0}

\begin{eqnarray}}{\end{eqnarray}\setcounter{equation}{\value{equationa}}}

\newcounter{example}
\newcounter{remark}
\newcounter{theorem}
\newcounter{proposition}
\newcounter{lemma}
\newcounter{corollary}
\newcounter{definition}

\def\thedefinition{\arabic{definition}}

\newcommand{\mar}[1]{}

\begin{document}
\hbox{}

{\parindent=0pt

{\large \bf On the notion of a differential operator in noncommutative
geometry}  
\bigskip 

{\bf G. Sardanashvily}

\medskip

\begin{small}

Department of Theoretical Physics, Moscow State University, 117234
Moscow, Russia

E-mail: sard@grav.phys.msu.su

URL: http://webcenter.ru/$\sim$sardan/
\bigskip

{\bf Abstract.}
The algebraic notion of a differential operator on a module over a
commutative ring is not extended to a module over a noncommutative
ring. 
\end{small}
}

\bigskip
\bigskip

Let $\cK$ be a commutative ring and $\cA$ a commutative $\cK$-ring. 
By a module $P$ over a commutative ring $\cA$ throughout is meant a
central bimodule, i.e., $ap=pa$ for all $p\in P$ and $a\in \cA$.
Let $Q$ be another $\cA$-module. 
There are three equivalent definitions of 
a $Q$-valued differential operator on an $\cA$-module $P$
\cite{vinb,book00,epr} (see items (i) -- (iii)). None of them is extended
to modules over a noncommutative ring. Therefore, a different
definition is suggested.

Note that the definition (iii) is based on the notion of 
jet modules of a commutative ring. These jet modules provide the
standard formulation of the Lagrangian formalism on modules over
commutative and graded commutative rings \cite{book00,lmp,mpl}.
However, the notion of a jet module fails to be extended to a
noncommutative ring. 

(i)
Let $P$ and $Q$ be modules over a commutative $\cK$-ring $\cA$.
The $\cK$-module $\hm_\cK (P,Q)$
of $\cK$-module homomorphisms $\f:P\to Q$ can be endowed with the
two $\cA$-module structures
\mar{5.29}\beq
(a\f)(p):= a\f(p),  \qquad  (a\star\f)(p): = \f (a p),\qquad a\in
\cA, \quad p\in P. \label{5.29}
\eeq
For the sake of convenience, we will refer to the second one as the
$\cA^\star$-module structure.
Let us put
\mar{spr172}\beq
\dl_c\f= c\f -c\star\f, \qquad c\in\cA. \label{spr172}
\eeq
An element $\Delta\in\hm_\cK(P,Q)$ is
called a $Q$-valued $s$-order differential operator on $P$ if
\be
\dl_{c_0}\circ\cdots\circ\dl_{c_s}\Delta=0
\ee
for any tuple of $s+1$ elements $c_0,\ldots,c_s$ of $\cA$.
The set $\dif_s(P,Q)$ of these operators
inherits the $\cA$-module structures 
(\ref{5.29}), and $\dif_{s-1}(P,Q)\subset \dif_s(P,Q)$. For
instance, zero order differential operators are $\cA$-module
homomorphisms of $P$ to $Q$, i.e., $\dif_0(P,Q)=\hm_\cA(P,Q)$. 
For our purpose, it suffices to consider first order differential operators. 
A first order differential operator
$\Delta$ satisfies the condition
\mar{n4}\beq
\dl_a\circ\dl_b\,\Delta(p)=\Delta(abp)-a\Delta(bp) -b\Delta(ap)+
ab\Delta(p)=0, \qquad b,a\in\cA.  \label{n4}
\eeq

(ii)
One can think of a ring $\cA$ as being an $\cA$-module generated by its unit
element $\bb$. 
A $Q$-valued first order differential operator
$\Delta$ on $\cA$ fulfils the relation
\mar{n10}\beq
\Delta(ab)=a\Delta(b)+ b\Delta(a)-ab \Delta(\bb), \qquad \forall a,b\in\cA.
\label{n10}
\eeq
It is a $Q$-valued derivation of $\cA$ if
$\Delta(\bb)=0$, i.e., the
Leibniz rule
\mar{+a20}\beq
\Delta(ab) = a\Delta(b) + b\Delta(a), \qquad \forall a,b\in \cA, \label{+a20}
\eeq
holds. Hence, any first order differential operator on $\cA$
falls into the sum  
\be
\Delta(a)= a\Delta(\bb) +[\Delta(a)-a\Delta(\bb)]
\ee
of a zero order differential operator $a\Delta(\bb)$ and a derivation
$\Delta(a)-a\Delta(\bb)$. 
Note that any zero order differential operator
$\Delta$ on $\cA$ 
is uniquely given by its 
value $\Delta(\bb)$.
Then, there is the $\cA$-module isomorphism $\dif_0(\cA,Q)=Q$ 
via the association
\be
Q\ni q\mapsto \Delta_q\in \dif_0(\cA,Q),
\ee
where $\Delta_q$ is defined by the equality
$\Delta_q(\bb)=q$. 
Let us consider the $\cA$-module morphism
\mar{n2}\beq
h: \dif_1(\cA,Q)\to Q, \qquad h(\Delta)=\Delta(\bb). \label{n2}
\eeq
One can show that any $Q$-valued first order differential operator 
$\Delta\in \dif_1(P,Q)$ on $P$ uniquely factorizes 
\mar{n13}\beq
\Delta:P\ar^{\gf_\Delta} \dif_1(\cA,Q)\ar^h Q \label{n13}
\eeq
through the morphism $h$ (\ref{n2}) 
and some homomorphism 
\mar{n0}\beq
\gf_\Delta: P\to \dif_1(\cA,Q), \qquad (\gf_\Delta p)(a)=\Delta(ap),
\qquad a\in \cA, \label{n0} 
\eeq
of the $\cA$-module $P$ to the $\cA^\star$-module $\dif_1(\cA,Q)$.
Hence, the assignment $\Delta\mapsto\gf_\Delta$ defines
the isomorphism 
\mar{n1}\beq
\dif_1(P,Q)=\hm_\cA(P,\dif_1(\cA,Q)). \label{n1}
\eeq

(iii)
Given an $\cA$-module $P$, let us consider the tensor product
$\cA\otimes_\cK P$ over $\cK$.
We put
\mar{spr173}\beq
\dl^b(a\otimes p)= (ba)\otimes p - a\otimes (b p), \qquad p\in
P, \qquad a,b\in\cA.  \label{spr173}
\eeq
Let us denote by $\m^{k+1}$ the submodule of $\cA\ot_\cK P$ generated 
by elements of the type
\be
\dl^{b_0}\circ \cdots \circ\dl^{b_k}(a\otimes p).
\ee
The $k$-order jet module $\gj^k(P)$ of the
module $P$ is defined as the quotient of the $\cK$-module $\cA\otimes_\cK
P$ by $\m^{k+1}$. We denote its elements $a\ot_kp$.
In particular,
the first order jet module $\gj^1(P)$
consists of elements $a\ot p$ modulo the relations
\mar{mos041}\beq
\dl^a\circ \dl^b(\bb\ot p)=\bb\ot(abp) -a\otimes
(bp) -b\otimes (ap) + ab\otimes p =0. \label{mos041}
\eeq
The $\cK$-module $\gj^1(P)$ is endowed with the following two $\cA$-module
structures:
\mar{+a21}\beq
b(a\ot_k p):= ba\ot_k p, \qquad
b\star(a\otimes_k p):= a\otimes_k (bp). \label{+a21}
\eeq
There exists the module morphism
\mar{5.44}\beq
J^1: P\ni p\to \bb\otimes_1 p\in \gj^1(P)
\label{5.44}
\eeq
of the $\cA$-module $P$ to the $\cA^\star$-module $\gj^1(P)$ such that
$\gj^1(P)$ seen as an $\cA$-module is generated by
elements $J^1p$, $p\in P$. Conversely, there is the epimorphism
\mar{+a13}\beq
\pi^1_0:\gj^1(P) \ni a\ot_1 p\to ap \in P.\label{+a13}
\eeq
A glance at $\dl_b$ (\ref{spr172}) and
$\dl^b$ (\ref{spr173}) shows that the morphism $J^1$ (\ref{5.44}) is a
first order differential operator on $P$. As a consequence,
any $Q$-valued first order differential operator $\Delta$ on 
$P$ uniquely factorizes
\mar{n20}\beq
\Delta: P\ar^{J^1} \gj^1(P)\ar^{\gf^\Delta} Q \label{n20}
\eeq
through the morphism $J^1$ (\ref{5.44}) and some $\cA$-module
homomorphism $\gf^\Delta: 
\gj^1(P)\to Q$.
The assignment $\Delta\mapsto {\got f}^\Delta$ defines 
the isomorphism
\mar{5.50}\beq
\dif_1(P,Q)=\hm_{\cA}(\gj^1(P),Q). \label{5.50}
\eeq

Now, let $\cA$ be a noncommutative $\cK$-ring and $P$ a $\cA$-bimodule.
Let $\cZ_\cA$ be the center of $\cA$ and $\cZ_P$ the center of $P$
(i.e., $\cZ_P$ consists of elements $p\in P$ such that $ap=pa$ for all
$a\in\cA$).  
Let $Q$ be another $\cA$-bimodule. The $\cK$-module $\hm_\cK(P,Q)$ can
be provided with the left $\cA$-module structures (\ref{5.29}) and the 
similar right ones. The left $\cA$-module homomorphisms $\Delta:P\to Q$
obey the conditions $\dl_c\Delta=0$ and, therefore, can be regarded as
left $Q$-valued zero order differential operators on $P$. One can also
write the condition (\ref{n4}). However, a problem is that, if $\cA$ is
noncommutative, zero order differential operators (e.g., $Q=P$ and
$\Delta=\id P$) fail to satisfy this condition.  

If $P=\cA$, a $Q$-valued zero order differential operator $\Delta$ on
$\cA$ takes its value $\Delta(\bb)$ only in the center $\cZ_Q$ of $Q$.
Therefore, one can rewrite the condition (\ref{n10}) as
\be
\Delta(ab)=a\Delta(b)+ \Delta(a)b -ab \Delta(\bb), \qquad \forall a,b\in\cA.
\ee
This provides the definition of a $Q$-valued first order differential
operator $\Delta$ on $\cA$. It is the sum of a $Q$-valued derivation $\dr$
of $\cA$ which obeys the Leibniz rule
\mar{n11}\beq
\dr(ab)=(\dr a)b +a\dr b, \qquad a,b\in\cA, \label{n11}
\eeq
and some zero order differential operator. Therefore, we have
$\Delta(\bb)\in \cZ_Q$.
The
$Q$-valued first order differential operators on $\cA$ make up a
left $\cZ_\cA$-module $\dif_1(\cA,Q)$. Then, one may try to define a 
$Q$-valued first order differential
operator on an $\cA$-bimodule $P$ as the composition
$\Delta=h\circ\gf$ (\ref{n13}). 
However, such an operator takes its values only in the center $\cZ_Q$ of $Q$
since $\Delta(p)=(\gf p)(\bb)\in \cZ_Q$.

The composition (\ref{n20}) fails to provide the definition
of a differential operator on a module over a noncommutative ring either.
The key point is that the jet module of a noncommutative ring is ill defined.
Namely, the projection $\pi^1_0$ (\ref{+a13}) of the zero element 
(\ref{mos041}) in $J^1(P)$ fails to be zero in $P$.

Using the fact that derivations of a
noncommutative $\cK$-ring $\cA$ with values in an $\cA$-bimodule are
well defined, one can suggest the following definition 
of first order differential operators on modules over a noncommutative rings.
Let $P$ and $Q$ be bimodules over a noncommutative $\cK$-ring $\cA$.
A $\cK$-module homomorphism $\Delta\in\hm_\cK(P,Q)$ of $P$ to $Q$ is
said to be a $Q$-valued first order differential operator on $P$ if it
obeys the condition
\mar{n21}\beq
\Delta(apb)=(\op\dr^\to a)(p)b +a\Delta(p)b + a(\op\dr^\leftarrow b)(p),
\label{n21}
\eeq
where $\op\dr^\to$ and $\op\dr^\leftarrow$ are derivations of $\cA$
which take their values in the modules $\hm_\cA^{\rm r}(P,Q)$ and 
$\hm_\cA^{\rm l}(P,Q)$ of right $\cA$-module homomorphisms and
left $\cA$-module homomorphisms of $P$ to $Q$, respectively. Namely, 
$(\op\dr^\to a)(pb)=(\op\dr^\to a)(p)b$ and $(\op\dr^\leftarrow b)(ap)
=a(\op\dr^\leftarrow b)(p)$. Note that $\hm_\cA^{\rm r}(P,Q)$ and
$\hm_\cA^{\rm l}(P,Q)$ are $\cA$-bimodules such that
\be
&& (a\f)(p):= a\f(p), \qquad (\f a)(p):= \f(ap), \qquad \f\in 
\hm_\cA^{\rm r}(P,Q),\\
&& (a\vf)(p):= \vf(pa), \qquad (\vf a)(p):= \vf(p)a, \qquad \vf\in 
\hm_\cA^{\rm l}(P,Q),
\ee
for all $a\in\cA$.

For instance, let $P=P^*$ be a differential calculus over a $\cK$-ring
$\cA$ provided with an associative multiplication $\circ$ and a
coboundary operator $d$. Then, $d$ is a $P$-valued first order
differential operator on $P$. It obeys the condition (\ref{n21}) which reads
\be
d(apb)=(da)\circ pb+a(dp)b + ap\circ db.
\ee

Another important example is a Dubois--Violette connection $\nabla$ on an
$\cA$-bimodule $P$ \cite{book00,epr,dub}. It associates to every
$\cA$-valued derivation $u$ of $\cA$ 
a $P$-valued first order differential operator $\nabla_u$ which
obeys the Leibniz rule
\be
\nabla_u(apb)=u(a)pb +a\nabla_u(p)b+apu(b).
\ee

\end{document}